\documentclass[preprint2]{aastex}

\newcommand{\etal}{et~al.\ }

\slugcomment{Galaxy Formation Forum}

\shorttitle{Star Formation of Early Type Galaxies}
\shortauthors{Schiavon}

\begin{document}


\title{History of Star Formation of Early Type Galaxies from Integrated
	Light: Clues from Stellar Ages and Abundances}

\author{Ricardo P. Schiavon}
\affil{Gemini Observatory, 670 N. A'ohoku Place, Hilo, HI 96720, USA}

\begin{abstract} 

I briefly review what has been recently learned from determinations of
mean stellar ages and abundances from integrated light
studies of early-type galaxies, and discuss some new questions posed
by recent data.  A short discussion of spectroscopic ages is
presented, but the main focus of this review is on the abundances
of Fe, Mg, Ca, N, and C, obtained from comparisons of measurements
taken in integrated spectra of galaxies with predictions from stellar
population synthesis models.  

\end{abstract}


\section{Introduction}

How galaxies form---and the stars within them---is one of the major
open questions of modern cosmology.  Early-type galaxies (ETGs)
host a large fraction of the stellar mass in today's universe, and
typically show no evidence for major ongoing star formation, so
they are natural targets for the investigation of how and when
galaxy assembly and star formation occurred in the past.  The task
is a difficult one, which has consumed a large amount of observational
and theoretical effort in the past several decades.  All that work
cannot be fairly summarized in a such short article.  This review is
thus narrowly focused on what has been learned from stellar ages
and abundances based on integrated light studies, and the new
questions raised by recent evidence.  Apologies go to many hard-working
colleagues for any important omissions.

\section{Stellar Ages}

Historically, the debate on the star formation history of ETGs was
framed in terms of two competing scenarios: hierarchical clustering
(e.g., White \& Rees 1978, Searle \& Zinn 1978), according to which
these galaxies were assembled through the merging of less massive
structures formed at high redshift; and monolithic dissipative
collapse (e.g., Eggen \etal 1962, Larson 1974), whereby massive
ETGs were formed at very high redshift by means of a rapid gravitational
collapse.  Deciding between these two scenarios was one of the main
motivations behind attempts to measure stellar ages and abundances
in ETGs.  Today there is little question that galaxies formed
hierarchically in a $\Lambda$-CDM universe, and while that historical
debate has been settled, studies of unresolved stellar populations
have acquired renewed importance, as they provide much needed
constraints to increasingly sophisticated galaxy formation models.

Early attempts at dating stellar populations from applications of
stellar population synthesis models to observations of integrated
light were based on photometric or low-resolution spectrophotometric
observations (e.g., O'Connell 1980, Gunn \etal 1981, Renzini \&
Buzzoni 1986) and high-resolution photographic spectroscopy (e.g.,
Rose 1985).  They led to promising, yet not entirely conclusive
results, due to limitations of the early models and/or uncertainties
associated with the age-metallicity degeneracy (e.g., Renzini 1986,
Worthey 1994).  The latter is a manifestation of the similar
dependence of the temperatures of main sequence and giant stars of
a given stellar population on age and metallicity, which causes the
integrated colors of stellar populations, particularly in the optical
and near-UV, to respond in similar ways to variations of these two
parameters.

It was only after the systematic modeling of Balmer lines as
(relatively) clean age indicators, initiated by Worthey (1994),
that reliable quantitative estimates of mean
luminosity-weighted\footnote{See Trager \& Somerville (2009) for a
discussion of the relation between luminosity- and mass-weighted
ages/abundances with those obtained from comparison of line indices with
single stellar population synthesis models.} stellar ages became
available.  The method relies on the dependence of Balmer lines,
such as H$\beta$ $\lambda$ 486 nm, on the temperatures and luminosities
of turnoff stars---which are higher in younger stellar populations.
For A-type stars and cooler, Balmer line strength is positively
correlated with temperature, so that the lines in integrated spectra
of stellar populations older than a few 100 Myr are stronger for
younger ages.  Spectroscopic ages based on $H\beta$ for large numbers
of ETGs, both in the field and in clusters, suggest that many of
them have undergone recent star formation events (e.g., Trager \etal
2000, Kuntschner 2000, Caldwell \etal 2003, Denicol\'o \etal 2005,
Thomas \etal 2005).  Trager \etal proposed a scenario where the
bulk of the stars in their sample galaxies were old, and only a
very small fraction of their stellar populations had young ages
($\sim$ 1 Gyr).  Because the latter are brighter in the optical,
they weight the mean ages towards lower values.  Although plausible,
this schematic scenario could not be verified by the Trager \etal
data because of the degeneracy between the age of the young component
and its relative contribution to a galaxy's total stellar mass
budget, which cannot be broken on the basis of $H\beta$ and metal-line
indices alone.

With the inclusion of additional age indicators, such as higher-order
Balmer lines in the blue (Leonardi \& Rose 1996, Worthey \& Ottaviani
1997, Schiavon 2007), stronger constraints were placed on the age
distributions of stars.  These studies confirmed early suggestions
that a small fraction of the stellar mass budget (few \%) is required
to be young in order to match the data.  They also ruled out claims
that Balmer line strengths in the spectra of ETGs required the
presence of metal-poor stars with blue horizontal branches (e.g.,
Freitas Pacheco \& Barbuy 1995, Maraston \& Thomas 2000, Lee \etal
2000).  Models including metal-poor stars cannot simultaneously
match all Balmer and metal lines in the 400-530 nm range (see also
Trager \etal 2005).

Perhaps the cleanest evidence for the presence of young/intermediate-age
stellar populations was provided by Galex observations of ETGs by
Yi \etal (2005).  Analyzing UV--optical colors of ETGs from the
Sloan Digital Sky Survey, Yi \etal found that approximately 2/3 of
their sample present strong UV fluxes which cannot be explained by
hot horizontal branch stars.  They estimate that approximately
1--2\% of the stellar mass is in the form of young stars.  Because
they constitute a tiny fraction of the stellar mass budget, young
stars are vastly outshone by older populations in the optical, where
they can only be detected on the basis of accurate spectrum synthesis
of very high S/N spectra (e.g., Schiavon et al.  2004, Schiavon
2007, Graves \etal 2007)---while being relatively easy to detect
in the UV.

More recent estimates based on a combination of Galex photometry
with larger SDSS samples indicate that as much as 10\% of all stellar
mass in ETGs today were formed since z $\sim$ 1 (S.\ Yi, this
volume).  While a fascinating debate is ongoing regarding what
triggered and quenched this relatively recent star formation (e.g.,
Kuntschner \etal 2010, Zhu \etal 2010, S\'anchez-Bl\'azquez \etal
2009, Serra \etal 2008, Schawinski \etal 2007, Kaviraj \etal 2007,
Graves \etal 2007), information on the history of formation of the
remaining $\sim$ 90\% of the stellar mass is relatively scanty,
possibly because it requires observations of large samples at higher
redshifts than so far possible and/or a far more thorough assessment
of the fossil record in the abundance patterns of stars in the
nearby ETGs.  The latter is the topic of the next section.

\section{Stellar Abundances}

Crucial information about the star formation history is encoded in
the chemical composition of stars.  The key observables are mean
(luminosity-weighted) metallicities, abundance ratios, and the run
of these quantities with $\sigma$.  It has long been known that the
central regions of nearby massive galaxies are metal-rich (e.g.,
Spinrad \& Taylor 1971).  It was only far more recently that reliable
metallicity estimates became possible for massive ETGs at cosmological
distances, showing that they have comparably high metallicities,
(e.g., Spinrad \etal 1997, J\o rgensen \etal 2005, Schiavon \etal
2006), implying a relatively rapid early chemical enrichment.  A
more detailed knowledge of the history of star formation requires
accurate estimates of detailed abundance patterns in a range of
redshifts, but work on abundance ratios has so far been restricted
to relatively nearby samples (but see J\o rgensen \etal 2005 and
Kelson \etal 2006).


\subsection{Magnesium}

The Mg I 517-518 nm doublet has been the chief metallicity indicator
in early studies of the chemical composition of ETGs from integrated
light, due to its strength and location in a spectral region where
astronomical detectors were very sensitive.  Early studies found
evidence for an overabundance of Mg relative to Fe in ETGs (e.g.,
Peterson 1976).  When Mg and Fe lines were finally compared with
models of stellar population synthesis, it was found that
[Mg/Fe]\footnote{\small More often than not, this result is phrased
in the literature in terms of an overabundance of $\alpha$ elements
relative to iron.  Despite the many theoretical reasons in favor
of the assumption that all $\alpha$ elements should vary {\it in
tandem}, there is so far no firm evidence that any $\alpha$ element
other than Mg is enhanced in ETGs, except perhaps for Ti (Milone
\etal 2000).} is above solar in the centers of massive ETGs (Worthey
\etal 1992).  This was arguably one of the most influential results
in the history of the field, and it has guided theoretical work to
this day.  Further studies showed that Mg enhancement is correlated
with central velocity dispersion ($\sigma$) and metallicity (e.g.,
J\o rgensen 1999, Trager \etal 2000, Thomas \etal 2005, Schiavon
2007, Smith \etal 2009).  At least three scenarios have been invoked
to explain this finding, all involving the balance between Mg
enrichment by SN II and Fe enrichment by SN Ia: short star formation
timescale, top-heavy IMF, and selective winds (Faber \etal 1992).
One popular interpretation of the data invokes the existence of a
relation between star formation timescale and galaxy mass (e.g.,
Thomas \etal 2005).  The data, however, did not allow one to rule
out scenarios based on selective winds or IMF variations.

Some exciting new results have been presented in a series of papers,
by G. Graves and collaborators.  Analyzing a large sample of stacked
SDSS\footnote{Sloan Digital Sky Survey} spectra, they mapped the
spectroscopic ages, [Fe/H], [Mg/H], and [Mg/Fe] of ETGs onto the
fundamental plane (FP) with particular attention to how these
parameters are distributed along and across the FP (Graves \etal
2009, Graves \& Faber 2010, Graves \etal 2010).  Besides recovering
the well known trends of age and abundances with $\sigma$, Graves
\etal showed that, at fixed $\sigma$, star formation histories of
ETGs correlate strongly with galaxy surface brightness.  Galaxies
with higher surface brightness have lower spectroscopic ages, higher
[Fe/H] and [Mg/H], but lower [Mg/Fe].  Graves \& Faber (2010) contend
that the thickness of the FP is accounted for by departures of a
constant dynamical-mass-to-light ratio ($M_{dyn}/L$), which may be
due to variations in either dark-matter fraction or IMF, and not
to effects due to passive evolution of the stellar populations.  At
fixed $M_{dyn}$, galaxies with higher surface brightness (located
"above" the FP) have a higher surface stellar density and therefore
are characterized by either a lower dark matter fraction or by a
bottom-heavier IMF.  They formed stars during a longer timescale,
so that their metallicities are higher, but both their spectroscopic
ages and [Mg/Fe]s are lower than their low-surface-brightness
counterparts.  Graves \etal propose a scenario where star formation
in galaxies with same $M_{dyn}$ was truncated at different times,
with longer/shorter star formation timescales resulting in higher/lower
stellar surface mass density and surface brightness, higher/lower
metallicity, lower/higher [Mg/Fe], and younger/older spectroscopic
ages.  In short, these new results single-handedly explain the
thickness of the FP and establish possible (and testable) correlations
between the star formation histories of ETGs and such measurable
quantities as dark-matter fraction and the shape of the low-mass
end of the stellar IMF.  Further progress will be determined by
observational tests of these predictions, as well as more sophisticated
chemodynamical modeling of ETG formation.



\subsection{Calcium}

Past studies suggest that Ca does not behave like Mg, with [Ca/Fe]
being possibly solar (or lower) and not correlated with $\sigma$.
Trager \etal (1998) found the Lick/IDS Ca4227 index to be essentially
independent of $\sigma$.  Accordingly, Thomas \etal (2003) concluded
that [Ca/Fe] in their sample galaxies was also essentially constant
with $\sigma$.  Saglia \etal (2002), on the other hand, found the
Ca II triplet (CaT, 849, 855, 862 nm) to be mildly decreasing with
$\sigma$.  Vazdekis \etal (2003) and Cenarro \etal (2004) compared
new single stellar population synthesis models for the CaT with data for field
and Coma galaxies, again finding very low [Ca/Fe].  While difficult
to understand, given that both Ca and Mg are $\alpha$ elements
manufactured in similar (though not identical) nucleosynthetic
sites, the implications of these results are potentially important,
giving theorists ample room for a wide range of speculations.

The unexpected behavior of Ca seems to be instead most likely caused
by difficulties in the interpretation of the measurements, particularly
because the two Ca indices employed in these studies do not respond
to Ca abundance variations in a clean fashion.  Prochaska \etal
(2005) showed that the Ca4227 index is severely affected by a CN
bandhead which contaminates the blue pseudocontinuum of the index,
making it lower.  Because CN is strongly correlated with $\sigma$
(Trager \etal 1998), the effect is stronger for higher $\sigma$ galaxies,
offsetting any dependence of the Ca line strength itself on $\sigma$,
thus making the index $\sigma$-independent.  Prochaska \etal
demonstrated this by defining a new index, Ca4227$_r$, which is
less affected by CN contamination.  They showed that Ca4227$_r$ is
as strongly correlated with $\sigma$ as Mg $b$.

While the issue of the slope of the Ca4227-$\sigma$ relation is
seemingly resolved, models that account for the effect of CN on the
Ca4227 index still indicate [Ca/Fe] $\sim$ 0 in massive ETGs (Schiavon
2007, Graves \etal 2007).  At face value, this confirms the abundance
ratios found by previous studies.  However, there may be non-negligible
systematics in the Ca abundances derived by application of the
Schiavon (2007) models.  They are affected by uncertainties in age,
and in the abundances of Fe, C, and N.  They are also affected by
uncertainties in the way models account for the contamination of
Ca4227 by CN.  So the matter should be considered far from settled.

Regarding the results based on CaT, one should bear in mind that
the integrated spectra of metal-rich stellar populations in the CaT
region is dominated by M giants (Schiavon \& Barbuy 1999), and that
fact has implications for both the zero point and the slope of the
[Ca/Fe]-$\sigma$ relation.  First let us consider the zero point.
The stellar libraries employed in the models used to analyze CaT
data in the past contain hardly any M giants with known metallicity,
let alone known [Ca/Fe] (Cenarro \etal 2001a).  Therefore, [Ca/Fe]
in the models themselves is uncertain, which obviously makes it
very hard for one to infer reliable [Ca/Fe] from comparison of those
models with the data.  As regards the slope of the CaT--$\sigma$
relation, we recall that the CaT lines are located in a region where
opacity in the spectra of M giants is dominated by TiO lines.  While
the definition of the CaT$^\star$ index employed in these studies
is partly meant to account for TiO contamination (Cenarro \etal
2001b), the index has not been shown to be immune to variations in
[Ti/Fe]\footnote{TiO lines are very sensitive to variations in the
abundance of Ti.} which may be important, given that there is
evidence that Ti is enhanced in ETGs (Milone \etal 2000).  Regarding
the negative slope of the CaT$^\star$ $\sim$ $\sigma$ relation,
that could be due to the effect of TiO opacity on the pseudocontinuum,
because: 1) TiO is well correlated with $\sigma$ (Trager \etal
1998), and 2) TiO lines are more sensitive to metallicity than CaT
lines (Schiavon \etal 2000, Schiavon \& Barbuy 1999, J\o rgensen
\etal 1992).  Finally, CN contamination of the CaT indices may also
be important (Erdelyi-Mendes \& Barbuy 1991).

In summary, we suggest that Ca abundances are far from well known
in ETGs, and there is no compelling motivation to resort to extreme
scenarios to account for the numbers currently available in the
literature.  More work is needed to produce reliable [Ca/Fe]
measurements in ETGs.

\subsection{Nitrogen \& Carbon} \label{cn}

While the behavior of C- and N-sensitive indices such as Lick CN$_1$,
CN$_2$, G4300 and C$_24668$ in ETG spectra has been well documented
for over a decade, it was only after the Schiavon (2007) models and
their implementation in EZ\_Ages (Graves \& Schiavon 2008) that
these indices could be intepreted in terms of [N/Fe] and [C/Fe]
(see also Kelson \etal 2006).  Both abundance ratios are found to
be super-solar and correlate strongly with $\sigma$ and metallicity
(Schiavon 2007, Graves \etal 2007, Smith \etal 2009).  This result
has been called into question recently by Toloba \etal (2009), who
found no correlation between the strength of the near-UV NH3360
feature and $\sigma$ in a sample of nearby galaxies.  They argue
that the NH3360 band is a clearer indicator of N abundance than the
Lick CN features used by EZ$\_$Ages, because the latter are also
dependent on C abundance.  The absence of a slope in the NH3360-$\sigma$
relation may be explained by the presence of metal-poor stars, whose
contribution to the integrated light is highest in the UV.  In fact,
multiple stellar population models show that the inclusion of a
small fraction of a metal-poor population flattens the NH3360-$\sigma$
relation, even in the presence of a [N/Fe]--$\sigma$ correlation
(G. Worthey, 2010, private communication).

The existence of a steep slope in the [N/Fe]--$\sigma$ relation,
if confirmed, is an important result, as it may indicate secondary
enrichment of N by stars ranging from 4--8 $M_\odot$ (Chiappini
\etal 2003).  Because these stars last for $\sim 10^8$ years, the
presence of a secondary-enrichment signature in the chemical
composition of stars in ETGs may constrain the lower limit for the
duration of star formation in the systems that formed the stars
that live today in those galaxies, (Schiavon 2007) and, perhaps
most importantly, their characteristic masses.  The increasing
evidence for the presence of multiple stellar populations in globular
clusters (Piotto 2009) may be an important clue in this regard.  It
has been long known that there is a marked spread in N and C
abundances in globular cluster stars (e.g., Smith \& Norris 1982,
Cannon \etal 1998, Carretta \etal 2005), which is roughly consistent
with enrichment by intermediate mass stars going through the AGB
phase (Ventura \& D'Antona 2008).  The presence of such CN
inhomogeneities seems to be a function of both cluster mass and
environment (Martell \& Smith 2009), as predicted by recent models
(Conroy \& Spergel 2010).

One may reasonably speculate that the evidence above indicates that
the stars we see in nearby ETGs were formed in the precursors of
today's Galactic globular clusters.  In that scenario, the signature
of secondary N enrichment we see in ETGs today would have been
established in those early systems, before they merged to form the
massive, dynamically hot galaxies we see today.  The Galactic halo
can be used as a resolved proxy to test this scenario.  The recent
identification of CN bimodality in a sample of halo field stars by
Martell \& Grebel (2010) argues for a similar process in operation
during the formation of the Galactic halo.  Because CN-strong stars
have almost certainly been formed in globular clusters (or their
precursors in the distant past), their presence in the halo field
is evidence of the early dissolution of those systems in the formation
of the Galactic halo.  Martell \& Grebel estimate that as much as
50\% of the halo mass may have been contributed by globular clusters
and their precursors.  If a similar process was responsible for the
assembly of stellar mass in ETGs, one might wonder whether a similar
fraction of the total mass would have been contributed by the
globular cluster precursors.  Could the slope of the [N/Fe] vs.
[Fe/H] or $\sigma$ be used to constrain that number?  What were the
characteristic masses of those systems?  Would it be possible to
construct chemodynamical evolution models for those low-mass systems
that are capable of reproducing all the abundance ratios measured
in today's ETGs?

Inclusion of C and N in abundance analyses of ETGs is bringing
interesting new insights on their star formation histories, which
could potentially even lead to a reinterpretation of the data on
[Mg/Fe].  Smith \etal (2009) analyzed a large data set for galaxies
from the Coma cluster and Shapley supercluster, spanning a very
wide range of $\sigma$.  They determined the abundances of several
elements using EZ\_Ages, then performed biparametric fits to the
relation between [Mg,Ca,C,N/Fe] and both [Fe/H] and $\sigma$, thus
disentangling the dependence of abundance ratios on these two
variables.  Smith \etal found that both [Mg/Fe] and [Ca/Fe] decrease
with [Fe/H], whereas [N/Fe] and [C/Fe] do not correlate with it.
They suggest that the run of [Mg/Fe] and [Ca/Fe] with [Fe/H] indicates
a short time scale for star formation.  They consider that the lack
of correlations of [N/Fe] and [C/Fe] with [Fe/H] is expected since,
unlike Mg and Ca, C and N are contributed by low(er) mass stars,
so that these elements should scale with Fe, not with Mg and Ca.
Interestingly, on the other hand, all abundance ratios show a strong
correlation with $\sigma$.  Smith \etal contend that this result
is difficult to interpret in terms of a simple dependence of star
formation timescale on galaxy mass (e.g., Thomas \etal 2005), because
that would preclude a correlation between [C/Fe] and [N/Fe] with
$\sigma$.  Clearly, more work is needed to clarify this issue.




\section{Concluding remarks}

The discussion above highlights the power of stellar population
synthesis to infer mean stellar properties from integrated light,
and constrain galaxy formation models.  Presently, well tested
stellar population synthesis models can be used to determine mean
stellar ages and abundances of Fe, Mg, Ca, C, and N.  The recent
addition of the latter two elements may spark the emergence of a
more complex picture of galaxy formation.  In order for that to
happen, more work is needed to refine chemodynamical models used
to interpret the abundance measurements and their relation with
global galaxy properties in terms of the physics of galaxy formation.
There has been very promising recent progress on this front, which,
due to space limitations, could not be reviewed here (e.g., Arrigoni
\etal 2010, Pipino \etal 2009a,b).  More theoretical work is also
needed to develop better stellar population synthesis models to
ascertain the reality of current abundance determinations, and to
include more elements in the pool of reliable abundances.  On this
front as well, different groups are making steady progress (Lee
\etal 2009, Coelho \etal 2007, Peterson 2007).  In that regard, it
is particularly desirable to extend stellar population synthesis
modelling towards the UV, in order to match the upcoming observing
capabilities that will make possible collection of large samples
of galaxy spectra at redshift beyond 2.  With the expected developments
in theory and observations, this field will likely go through very
exciting times in the next decade.


\acknowledgments

The author thanks Yonsei University and the organizers, especially
Suk-Jin Yoon and Sukyoung Yi, for a truly delightful workshop.  The
hospitality of the Department of Astrophysical Sciences at Princeton
University, where this paper was partly conceived, is warmly
acknowledged.  Conversations with David Spergel, Charlie Conroy,
Jenny Graves, Inger J\o rgensen, and Richard McDermid contributed
substantially to the formulation of some of the ideas presented in
this paper.  This work was supported by Gemini Observatory, which
is operated by the Association of Universities for Research in
Astronomy, Inc., on behalf of the international Gemini partnership
of Argentina, Australia, Brazil, Canada, Chile, the United Kingdom,
and the United States of America.

{}
\clearpage

\end{document}